\documentclass[a4paper,11pt]{article}
\usepackage{jheppub} 

\usepackage{xcolor}

\title{\boldmath Momentum Analyticity of Transverse Polarization Tensor  in the Normal Phase of a Holographic Superconductor }

\author[a,c]{Lei Yin $\star$} 
\author[b,c]{Hai-cang Ren $\ast$} 
\author[a]{Ting Kuo Lee $\dagger$} 
\author[c]{Defu Hou $\ddagger$}

\affiliation[a]{Institute of Physics, Academica Sinica ,\\Taipei   11529, P.O.C.} 
\affiliation[b]{Physics Department, The Rockefeller University,  \\ 1230 York Avenue, New York, 10021-6399, U.S.A.} %
\affiliation[c]{Institute of Particle Physics and Key Laboratory of Quark and Lepton Physics (MOS) , Central China Normal University, \\ Wuhan,  430079, P.R.O.C.} %

\emailAdd{$\star$ leiyinbox@gmail.com} 
\emailAdd{$\ast$ ren@mail.rockefeller.edu} 
\emailAdd{$\dagger$ tklee@phys.sinica.edu.tw} 
\emailAdd{ $\ddagger$ houdf@mail.ccnu.edu.cn}

\abstract{ 
  We explore the momentum analyticity of the static transverse polarization tensor of a 2+1 dimensional holographic superconductor in its normal phase, aiming at finding the holographic counterpart of the singularities underlying the Friedel oscillations of an ordinary field theory.  We prove that the polarization tensor is a meromorphic function with an infinite number of poles located on the complex momentum plane off real axis. With the aid of the WKB approximation these poles are found to lies asymptotically along two straight lines parallel to the imaginary axis for a large momentum magnitude. The similarity between the holographic Green’s function and that of an weakly coupled ordinary field theory (e.g., 2+1 dimensional QED) regarding the location of the momentum singularities offers further support to the validity of the gauge/gravity duality.
}

 
\begin{document}  

\maketitle 
\flushbottom 

\section{Introduction} 
\label{sec-1} 
The strongly correlated systems, such as QCD at low energy and cuprate superconductors remain major challenges in high energy physics and condensed matter physics. The perturbative expansion becomes unreliable in strong coupling and the first principle numerical simulation is often hindered by the fermion sign problem. The gauge/gravity duality \cite{Witten1998b,Maldacena1998,Aharony1999,Klebanov1999} opens a new avenue towards a qualitative or even quantitative understanding of some universal mechanisms, if any, behind the intriguing experimental observations in these systems \cite{Maldacena1998b,Policastro2001} . The holographic superconductor 
provides such an example, where the quantum effective action of a strongly coupled superconductor in 2+1 dimensions corresponds to a 3+1 dimensional classical action of an Abelian-Higgs theory coupled to the gravity with a AdS-Reissner-Nordst\"{o}m (AdS-RN) black hole \cite{Gubser2008}, or with a AdS- Schwarzschild black hole  in the probe limit \cite{Hartnoll2008,Hartnoll2008a}. The implied ratio between the AC conductivity threshold and the critical temperature comes close to the observed values from cuprates.

All holographic models used in condensed matter physics follow a bottom-up approach without the knowledge of the explicit Lagrangian underlying the quantum effective action implied by 
the gauge/gravity duality, as is reflected, for instance, in the lack of the link between the order parameter in the holographic superconductivity and the Cooper pairing of the fermionic degrees of freedom. Therefore in addition to expand the horizon of the holographic models, it is equally important to examine if the quantum effective action 
implied by holography possesses all fundamental properties of an ordinary field theory. The work reported below serves the latter purpose and addresses the analyticity of the Green's 
function with respect to the momentum. There have been extensive researches concerning the analyticity of the holographic Green's functions on the complex energy plane \cite{Son2002,Policastro2002} , the analyticity on the 
momentum plane was less explored and we would like to fill this gap.

Unlike the analyticity on the complex energy plane which is dictated by the unitary of the underlying field theory, the physics foundation of the analyticity on the momentum plane is less 
transparent. For a thermal field theory with Lorentz invariance, a new field theory can be constructed by interchange the roles of the energy and one component of momentum and 
the unitarity of the new theory dictates the momentum plane singularities of the original theory along the imaginary axis and gives rise to the exponential decay of the correlation 
functions in coordinate representation. It was shown in \cite{Hou2010} that this property was carried over to the Green functions  extracted from the AdS/CFT correspondence at zero chemical potential. Turning on the chemical potential, Lorentz symmetry is explicitly broken and there is not a priori argument regarding the analyticity on the complex 
momentum plane. The charge fluctuation in a Fermi liquid displays Friedel oscillations in coordinate space, which may be responsible to the superconductivity mechanism in some strongly correlated electronic system \cite{Kohn1965,Galitski2003}.  
This oscillation is attributed to the singularities of the (00)-component of the polarization tensor off-imaginary axis of the momentum plane and can be located explicitly at one-loop 
order. These complex singularities are also shared by other components of the polarization tensor and remain beyond the perturbation, say in Luttinger liquid \cite{Egger1995} . Locating 
such singularities of the holographic polarization tensor is the main theme of the present paper.

This research was motivated in part by the recent work \cite{Blake2015} where the complex momentum singularities was found numerically for the normal phase of a 2+1 dimensional 
holographic superconductor. We shall provide an analytic proof of their existence. It follows from the method in \cite{Hou2010} , the momentum singularities of the 
polarization tensor remain poles along the imaginary axis in the probe limit which was mostly investigated in literature. Going beyond the probe limit, one has to find the solutions 
of the full linearized Einstein-Maxwell equations in the RN blackhole background and the analytic approach becomes rather complicated. Aligning the momentum in x-direction, 
the fluctuations of the metric and the gauge potential can be divided to two decoupled groups according to the parity under reflection $y\to -y$. The odd parity group contains less 
number of components \cite{Edalati2010c} , $(a_y, h_{ty}, h_{xy})$, and will be explored below and the more complicated even parity group, will be reported elsewhere. We shall prove that similar to the
zero chemical potential case in \cite{Hou2010}, the polarization tensor at a nonzero chemical potential remains a meromorphic  function of the momentum, but with poles distributed asymptotically along two lines parallel to the imaginary axis at large momentum.

This paper is organized as follows. The momentum analyticity of the polarization tensor to one-loop order will be discussed in the next section with the result as a benchmark for comparison
with its holographic counterpart. In section 3, we shall review the holographic formulation in literature and provide a rigorous proof that the polarization tensor of the holographic 
superconductor in the normal phase is a meromorphic function of its momentum. The location of the complex momentum poles at large momentum will be explored in section 4 with the aid of the 
WKB approximation. Section 5 will conclude the paper.

\section{Photon Self-energy in $D =2+1$ Dimensional  Thermal QED}
\label{sec-2}

In a $2+1$ dimensional spacetime, the three $\gamma$ matrices  can be chosen as follows
\begin{align}
  \gamma_0 = \sigma_3 , \quad \gamma_1 = - \mathrm{i} \sigma_1, \quad \gamma_2 = - \mathrm{i}  \sigma_2
\end{align}
where $\sigma_i$ are the Pauli matrices. The Lagrangian density of a massless fermion field in an external electromagnetic potential reads  
\begin{equation}
\mathcal{L}[\psi,\bar\psi]=-\bar\psi\gamma_\lambda\left(\frac{\partial}{\partial x_\lambda}-ieA_\lambda\right)\psi+\mu\bar\psi\sigma_3\psi
\label{lagrange}
\end{equation}
where $e$ is the 2+1 dimensional electric charge. The chemical potential $\mu$ is included in (\ref{lagrange}) so that the grand partition function of the system at a temperature 
$T=1/\beta$ is given by the following path integral. 
\begin{align}
  \mathcal{Z}_\text{Fermion} = \int_{\psi_\alpha(\beta) = -\psi_\alpha(0)} \mathcal{D}\big(\psi_\alpha^* , \psi_\alpha\big) \, \exp\bigg\{ - \int_0^\beta d\tau \int d^3\vec r
\mathcal{L}[\psi,\bar\psi] \bigg\}
  \label{eq:2}  
\end{align}
The free fermion propagator in energy-momentum space takes the form 
\begin{align} 
 S_F (p) = \frac{\mathrm{i}}{(\mathrm{i} \nu + \mu) \sigma_3 - \mathrm{i} \vec{\sigma} \cdot \vec{p}}  
\label{eq:1}
\end{align}
where $\vec \sigma = (\sigma_1 , \sigma_2)$ and the Matsubara energy $\nu = 2 \pi T(n+\frac{1}{2}), \quad n \in \mathbb{Z}$.
To explore the transverse polarization in the static limit, we align the photon momentum along x-direction and evaluate the $yy$ component 
of the polarization tensor, $\Pi_{\mu\nu}(q)$, specified by the one loop Feynman diagram in Fig.~\ref{fig:one-loop}:

\begin{figure}[!htb]
  \centering{\includegraphics[width=3.5cm]{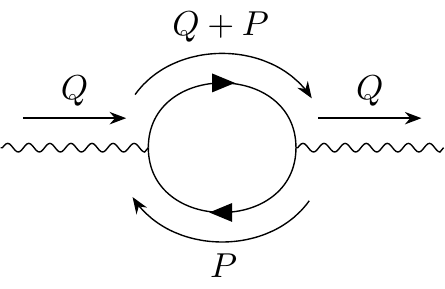}}
\caption{ the one-loop diagram in hot QED}
  \label{fig:one-loop}
\end{figure}
We have 
\begin{align}
  \Pi_{yy} (q)\equiv e^2 \sigma^\mathrm{tr}(q)
\end{align}
where $\sigma^\mathrm{tr}(q)$ is given by
\begin{align}
  \sigma^\mathrm{tr} (q) = \frac{1}{\beta} \sum_{n=-\infty}^{+\infty} \int \frac{\mathrm{d}^2 \vec p}{(2\pi)^2} \mathrm{Tr}\,  \bigg[ \sigma_2 S_F(P) \sigma_2 S_F(P+Q) \bigg]
\label{eq:4}
\end{align}
with $P = (\mathrm{i} \nu , \vec p) ; \, Q=(0,  \vec q)$ and $\vec q =(q,0)$.

Evaluating the trace in (\ref{eq:4}) and converting the Matsubara frequency sum there into a contour integral, we obtain:
\begin{equation}
\begin{aligned}
 \sigma^\text{tr}(q) &= -2 \int \frac{\mathrm{d}^2 \vec p}{(2 \pi)^2} \oint_{\mathcal{C}_1} \frac{\mathrm{d} z}{2 \pi \mathrm{i}} \; \frac{1}{e^{\beta z } +1} \frac{-(z+\mu)^2 + (p_1 + q) p_1 - p_2^2}{ \big[(z+\mu)^2 - |\vec p + \vec q|^2 \big]\, \big[(z + \mu)^2 - p^2\big]}  \\
 &= - \int \frac{\mathrm{d}^2 \vec p}{(2 \pi)^2} \, \bigg\{ \left[\frac{1}{e^{\beta(| \vec p + \vec q| -\mu)} +1}-\frac{1}{e^{-\beta(|\vec p + \vec q| +\mu)} +1}\right]
     \frac{-(\vec p + \vec q)^2 + (p_1 + q) p_1 -p_2^2}{|\vec p + \vec q| \big[ (\vec p + \vec q)^2 - \vec p^2 \big]}     \label{eq:5}  \\
 & \hspace{2.7cm}- \left[\frac{1}{ e^{\beta (p - \mu)} + 1}-\frac{1}{ e^{-\beta (p + \mu)} + 1}\right] \frac{-p^2 + (p_1 + q)p_1 - p_2^2}{p \big[ (\vec p + \vec q)^2 - p^2\big]} \bigg\}    \\
&= \sigma_\text{vac}^\text{tr}(q) +  \sigma_\text{matt}^\text{tr}(q) 
\end{aligned}
\end{equation}
where  $\sigma_\text{vac}^\text{tr}(q) = \sigma^\text{tr}(q)\, \bigg|_{T=0; \, \mu=0}$ and
\begin{equation}
\begin{aligned}
 \sigma_\text{matt}^\text{tr}(q)   &= - \int \frac{\mathrm{d}^2 \vec p}{(2 \pi)^2} \, \bigg\{ \frac{1}{e^{\beta(| \vec p + \vec q| -\mu)} +1} \frac{-(\vec p + \vec q)^2 + (p_1 + q) p_1 -p_2^2}{|\vec p + \vec q| \big[ (\vec p + \vec q)^2 - \vec p^2 \big]}     \label{eq:5-0}  \\
 & \hspace{2.7cm}- \frac{1}{ e^{\beta (p - \mu)} + 1} \frac{-p^2 + (p_1 + q)p_1 - p_2^2}{p \big[ (\vec p + \vec q)^2 - p^2\big]} \bigg\}  + (\mu \leftrightarrow -\mu)  \\
 \end{aligned}
\end{equation}
The convergence of the integral relies on the cancellation of the singularity at $(\vec p + \vec q)^2 - p^2=0$ in the sum of the two terms inside the curly bracket. Adding an infinitesimal imaginary part to the factor $(\vec p + \vec q)^2 - p^2$ 
in the denominators renders the integral of each term convergent without modifying the real part of the result. Then the integration momentum of the first term can be shifted to remove 
the $q$ dependence of the Fermi distribution function and we arrive at
\begin{equation}
\begin{aligned}
  \sigma^\text{tr}_\text{matt}(q) &= - \mathrm{Re} \; \int  \frac{\mathrm{d}^2 \vec p}{(2 \pi)^2} \, \bigg\{ \frac{1}{e^{\beta(p - \mu)} + 1} \times \frac{1}{p} \times  \\
 & \hspace{1.2cm}  \left[ \frac{-p^2 + p_1(p_1 - q) - p_2^2}{p^2 - (\vec p - \vec q)^2 + \mathrm{i}0^+}  - \frac{-p^2 + (p_1 + q) p_1 - p_2^2}{(\vec p + \vec q)^2 - p^2 + \mathrm{i}0^+}\right]  + (\mu \leftrightarrow -\mu) \bigg\}    \\
 &= - \frac{2}{\pi q} \int_0^{\frac{q}{2}} \ \mathrm{d}p \; \frac{p^2}{\sqrt{q^2 - 4 p^2}} \left[ \frac{1}{e^{\beta(p - \mu)} +1} + \frac{1}{e^{\beta(p + \mu)} + 1} \right]   \label{eq:7}
\end{aligned}
\end{equation}
where we have integrated out the polar angle of $\vec p$.

    To explore the analytic continuation of (\ref{eq:7}) to the complex $q$-plane, we employ the identity (\ref{eq:8}) and turn (\ref{eq:7}) into the following form
\begin{equation}
\begin{aligned}
  \sigma^\text{tr}_\text{matt}(q) &= - \frac{2}{\pi q} \int_0^{\frac{q}{2}} \ \mathrm{d}p \, \frac{p^2}{\sqrt{q^2 - 4p^2}} \left[ 1 + T \sum_\nu \left( \frac{1}{\mathrm{i} \nu + \mu - p } - \frac{1}{\mathrm{i} \nu + \mu + p}\right) \right]  \\
   &= -\frac{q}{16}-\frac{qT}{4\pi} \sum_\nu \int_0^1 \ \mathrm{d}x \, \frac{x^2}{\sqrt{1 - x^2}}  
\left( \frac{1}{\mathrm{i}\nu + \mu - \frac{q}{2}x } - \frac{1}{\mathrm{i}\nu + \mu +  \frac{q}{2}x}\right)     \label{series}
\end{aligned}
\end{equation}
which displays explicitly the complex singularities at $q = \pm 2 ( \mathrm{i} \nu + \mu)$ , here $\nu = \pi T(2n+1)$ and $n$ is an integer. To find out the nature of each singularity, we let $q$ approach to one of them, i.e.
\begin{align}
  \frac{q}{2} =  \mu + \mathrm{i}\nu    + \epsilon
  \label{eq:11}
\end{align}
with  $|\epsilon| \to 0$,  only the term with Matsubara energy $\nu$ in the series of (\ref{series}) contributes to the singularity and (\ref{series}) is switched into 
\begin{equation}
\begin{aligned}
  \sigma^\text{tr}(q) &= - \frac{T q}{4 \pi} \bigg[ \int_{-1}^1 \frac{\mathrm{d}z}{\sqrt{1 - z^2} \big( [\mu + \mathrm{i} \nu] (1 - z) - \epsilon z \big)} \\ 
    &  \hspace{3cm}  - \int_{-1}^1 \  \mathrm{d}z  \; \frac{\sqrt{1 - z^2}}{[\mu + \mathrm{i}\nu] (1 - z ) - \epsilon z} \bigg]+\hbox{non-singular terms}  \label{eq:12}
\end{aligned}
\end{equation}
where the two integrals can be calculated by means of residue theorem with the contour like Fig. \ref{fig:Contour} :
\begin{figure} [!htb]
  \centering{\includegraphics[width=3.5cm]{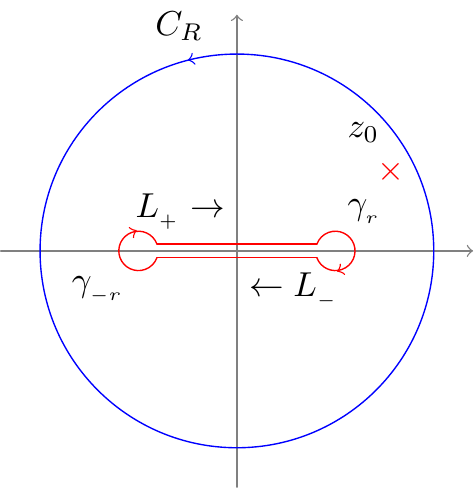}}
\caption{ The contour encloses the pole $z_0 := \frac{\mu + \mathrm{i} \nu}{\mu + \mathrm{i}\mu + \epsilon}$ , the red part is the branch cut  }\label{fig:Contour}
\end{figure}
Finally, the result of (\ref{eq:12}) is obtained:

\begin{align}
  \sigma^\text{tr}(q) = \mathrm{i} \frac{T q}{4} \frac{1}{ \sqrt{ 2 \epsilon (\mu + \mathrm{i}\nu)}} + \text{non-singular terms} 
 \label{eq:15}
\end{align}
If we set   
\begin{align}
  \frac{q}{2} = -[\mu + \mathrm{i}\nu] + \epsilon
\end{align}
here $|\epsilon| \to 0$,  we also can obtain a result similar to (\ref{eq:15}). Therefore,  the divergent behavior of $\sigma^\text{tr}(q)$ is like $O\big(\dfrac{1}{\sqrt{\epsilon}} \big)$ , and the singular points 
\begin{align}
  q = \pm 2 (\mu + \mathrm{i} \nu)
\end{align}
are square root branch points of the transverse polarization. These singularities generates the Friedel oscillations in the coordinate representation of $\sigma^\text{tr}(q)$.  In higher dimension, one can obtain a similar consequence, the purpose we study 2+1 dimensional field theory here is to compare to a case in gauge/gravity duality.

\section{Static Transverse Correlation in  Gauge/Gravity   Duality}
\label{sec-3}

\subsection{Gravity Preliminaries in the Bulk Theory}
\label{sec-3.1}

The  Gauge/Gravity correspondence employed in this work is encoded in the following relationship between a classical gravity-matter system in the bulk spacetime and a strongly coupled 
quantum field theory at the asymptotically AdS boundary.
\begin{align}
  \Gamma[\phi_0] = S_\mathrm{cl.}[\phi_0] \label{eq:16}
\end{align}
where $\phi_0$ denotes the collection of boundary values of bulk fields (gauge field, metric fluctuations, etc.), $\Gamma[\phi_0]$ is the quantum effective action (the generating functional
of proper vertices of the boundary field theory in strong coupling) and the classical bulk action $S_\mathrm{cl.}[\phi_0]$ is on-shell, i.e., evaluated at the solution of equations of motion subject to 
the boundary conditions specified by $\phi_0$.
We shall use the relation (\ref{eq:16}) to investigate the correlation functions of a $U(1)$ gauge at the boundary via calculating the on-shell action in the bulk. 
It's expected that the holographic method may shed some light on the conundrum on strongly-coupled physical system. 

Being parallel to the  QED case in Sec. 2, we are interested in the transverse polarization of the $U(1)$ gauge field in strong coupling, which is coupled to the metric fluctuations 
around a Reissner - Nordstr\"{o}m  black hole in an asymptotically Anti-de Sitter  spacetime (RN-AdS spacetime for short) .  
The bulk gravity-matter system, from which the holographic polarization tensor is extracted, is described by the Einstein-Maxwell action
\begin{align}
  S &= \int \ \mathrm{d}^D x \, \sqrt{-g}\; \bigg\{   G_D\, \big( R - 2 \Lambda \big)-  K_D\, \big( F_{\mu \nu} F^{\mu \nu} \big)  \bigg\}  \label{eq:17}
\end{align}
where $R$ is the scalar curvature, $\Lambda$ is the negative cosmological constant, $\Lambda = -\frac{1}{2}\frac{(D-1)(D-2)}{L^2}$,  $L$ is the AdS radius, and $K_D$ is the 
coupling constant in $D=d+1$ dimensional spacetime. This action at $d=3$ corresponds to the normal phase of the 2-dimensional  holographic superconductivity, where the scalar field vanishes,  investigated extensively in literature.  Here we took $\frac{K_4}{G_4} = L^2$. 

The presence of the $2$-form field $F_{\mu\nu} = \partial_\mu A_\nu - \partial_\nu A_\mu$ leads to a solution of equations of motion which consists of a 
charged RN-AdS black hole, given by the metric and a background gauge potential. We choose the coordinate system such that the black hole metric is given by
\begin{align}
  \ \mathrm{d}\bar s^2 = \bar g_{\mu \nu} \ \mathrm{d}x^\mu \ \mathrm{d}x^\nu = \frac{L^2}{z^2} \left( -f(z) \ \mathrm{d}t^2 + \frac{\mathrm{d}z^2}{f(z)} 
+ \ \mathrm{d}x^2 + \ \mathrm{d}y^2 \right)  \label{eq:18}
\end{align}
with
\begin{align}
  f(z) = 1 - (1+Q^2) \, \left(\frac{z}{z_+} \right)^3 +  Q^2 \, \left(\frac{z}{z_+} \right)^4   \label{eq:19} 
\end{align}
and the background gauge potential by
\begin{align}
  \bar A = \bar A_t \ \mathrm{d}t = \mu \left( 1 - \frac{z}{z_+}\right) \ \mathrm{d}t   \label{eq:20}
\end{align}
where $z_+$ is the coordinate of the horizon, $Q$ is the charge of the black hole and $\mu$ corresponds to the chemical potential of the boundary field theory.
The black hole charge is related to the chemical potential via
\begin{align}
  Q = \mu z_+    \label{eq:21}
\end{align}
and the Hawking temperature 
\begin{align}
  T = \frac{1}{4 \pi z_+}\left(  3 - Q^2 \right)=\frac{\mu}{4\pi }\frac{3-Q^2}{Q} \label{eq:10}
\end{align}
which is also the temperature of the boundary field theory. The positivity of the temperature requires $Q^2<3$, which makes the location of the second horizon 
of the RN-AdS black hole at $z>z_+$. The asymptotic AdS boundary is located at $z=0$ and the physical domain of the radial coordinate is thereby $0\le z\le z_+$. 
The solution (\ref{eq:18}) and (\ref{eq:20}) define a thermal bath of the boundary field theory. 

\subsection{Fluctuations of Gauge Field and Metric Field}
\label{sec:fluct-gauge-field}

According to the holographic principle, the electric current operator and the energy-momentum tensor on the boundary field theory are dual to the fluctuations of the 
gauge field and metric field in the bulk theory respectively. In contrast to the perturbation theory, the retarded Green functions on 
boundary theory extracted from the duality corresponds to the strong coupling limit of them in the boundary field theory. 
Our interest here is in the static transverse Green function.

To be begin with, we define 
\begin{equation}
\begin{aligned}
  g_{\mu \nu} &= \bar g_{\mu \nu} + h_{\mu \nu}   \\
  A_\mu  &= \bar A_\mu + a_\mu     \label{eq:22}
\end{aligned}
\end{equation}
where ($\bar g_{\mu \nu} , \bar A_\mu$) are the background fields, and ($h_{\mu \nu}, a_\mu$) are the fluctuations respectively. To maintain  the basic property of a metric 
tensor $ g_{\mu \rho} g^{\rho \nu}=\delta_\mu^\nu $ , we have
\begin{align}
 g^{\mu \nu} = \bar g^{\mu \nu} - h^{\mu\nu} , \quad  \text{and} \quad \sqrt{-g} = \sqrt{ -\bar g } \big( 1 + \frac{1}{2}\bar g^{\mu \nu} h_{\mu \nu} + O\big( h^2 \big) \big)
\end{align}
To extract two-point Green's functions, we need only the linearized Einstein-Maxwell equations in fluctuations, all dynamic quantities are considered up to the 1st-order in 
$h_{\mu \nu}$ and $a_\mu$. e.g. $h_{\;t}^y \equiv g^{yy} h_{ty} = \bar g^{yy} h_{ty}$. The fluctuations with energy and momentum $(\omega, q)$ take the form
\begin{equation}
\begin{aligned}
  h_{\mu \nu} (t, z, x ,y) &\sim e^{ \mathrm{i}(-\omega t + q x) } \, h_{\mu \nu} (z | \omega, q)  \\
a_\mu(t, z , x, y) &\sim e^{\mathrm{i}(-\omega t + q x)} \, a_\mu(z | \omega, q)   \label{eq:24}
\end{aligned}
\end{equation}
where we have aligned the momentum along the x-axis by taking advantages of the $SO(2)$ symmetry in $x-y$ plane. 
We work in the radial gauge
\begin{align}
  h_{z \nu} = 0 , \qquad a_z = 0  \label{eq:23}
\end{align}
throghout the paper where
the fluctuation fields are classified into two categories according to the parity under $y \to -y$
\begin{align}
  \text{Odd Parity:} &\quad h_{ty} , \; h_{xy} , \;  a_y \\
  \text{Even Parity:} &\quad h_{tt} , \; h_{tx} , \;  h_{xx} , \; h_{yy} ,\; a_t ,\; a_x 
\end{align}
At a nonzero temperature, the linearized Einstein-Maxwell equations for $h_{\mu \nu} (z | \omega, q)$ and $a_\mu(z | \omega, q)$ are
\begin{align}
  f\big[{h_{\; t}^y}''  - \frac{2}{u} {h_{\; t}^y}' -  \frac{ 4 Q^2}{\mu} \, u^2 {a_y}' \big] -Q^2 \, \mathfrak{w} \mathfrak{q} h_{\; y}^x - Q^2 \, \mathfrak{q}^2 h_{\; t}^y  &= 0    \label{eq:25} \\
  f\big[ f {h_{\; y}^x}'' + (f' - \frac{2f}{u}) {h_{\; y}^x}' \big] + Q^2 \, \mathfrak{w}^2 h_{\; y}^x + Q^2 \, \mathfrak{w q} \, h_{\; t}^y   &= 0  \label{eq:26}\\
f \big[ f {a_y}'' + f' {a_y}' - \mu \, {h_{\; t}^y}' \big] + Q^2 (\mathfrak{w}^2 - f \mathfrak{q}^2) \, a_y &= 0  \label{eq:27}  \\
f \, \mathfrak{q} {h_{\; y}^x}' + \mathfrak{w} \, {h_{\; t}^y}' - 4z_+ Q \, \mathfrak{w} u^2 \, a_y &= 0  \label{eq:28}
\end{align}
where we define three dimensionless quantities. 
\begin{align}
  \mathfrak{w} = \frac{\omega}{\mu}, \quad \mathfrak{q} = \frac{q}{\mu} , \quad u = \frac{z}{z_+} \label{scaling}  
\end{align}
and set the AdS radius $L = 1$. The prime refers to the derivative with respect to $u$ and $f=1-(1+Q^2)u^3+Q^2u^4$. Among the four equations (\ref{eq:25})-(\ref{eq:28}), only three of them are independent.

 In the static limit, $\omega =0$, $h^x_y$ decouples from $h^y_t$ and $a_y$ and we are left with two independent equations  
\begin{align}
  f \big[f {a_y}'' + f' {a_y}' - \mu {h_{\; t}^y}' \big] - Q^2 \, f \mathfrak{q}^2\, a_y &= 0 \label{eq:29} \\
f \big[ {h_{\; t}^y}'' - \frac{2}{u} \, {h_{\; t}^y}' - \frac{4 Q^2}{\mu} u^2 \, {a_y}' \big] - Q^2 \mathfrak{q}^2 \, h_{\; t}^y &= 0 \label{eq:30}
\end{align} 
The solution of (\ref{eq:29}) and (\ref{eq:30}) can be obtained in terms of a pair of auxiliary fields\cite{Edalati2010c}, the master fields $\Phi_\pm$, which satisfy a pair of 
decoupled differential equations\cite{Kodama2004}. We have
\begin{align}
  \Phi_\pm(u) = - \frac{\mu}{u} {h_{\; t}^y}'(u) + 2 Q^2 \big[ 2u - g_\pm(k) \big] \, a_y(u)   \label{eq:31}
\end{align}
where   
\begin{align}
  g_\pm (k) &:= \frac{3}{4}\left(1 + \frac{1}{Q^2} \right) \pm k \label{eq:46}  
\end{align}
and the modified momentum $k$ is related to the dimensionless momentum $\mathfrak{q}$ via
\begin{align}
  k^2 &:= \mathfrak{q}^2 + \left[ \frac{3}{4}\left( 1 + \frac{1}{Q^2}\right)\right]^2 \label{eq:47}
\end{align}
The differential equations satisfied by the master fields in the static limit read
\begin{align}
  \Phi_\pm'' + \frac{f'}{f}\Phi_\pm' + \left[ -\frac{f'}{uf} - \frac{\mathfrak{q}^2 Q^2}{f}  - \frac{2 Q^2 u}{f} g_\pm(k) \right] \, \Phi_\pm = 0  \label{eq:32}
\end{align}
Each of them is a Fuchs equation with five regular points (four roots of $f(u)$ and the infinity). Upon a transformation
\begin{align}
  \Phi_\pm = \frac{1}{\sqrt{f}}\, \Psi_\pm  \label{eq:35}
\end{align}
The differential equation satisfied by each of $\Psi_\pm$,
\begin{align}
  \Psi_\pm''(u | \mathfrak{q}) + V_\pm(u | \mathfrak{q}) \, \Psi_\pm(u | \mathfrak{q})  =  0 \label{eq:36}  
\end{align}
is of Schr\"{o}dinger type with the potential energy
\begin{align}
  V_\pm(u | \mathfrak{q}) =  - \frac{ Q^2}{f} \mathfrak{q}^2  - \frac{2 Q^2 u}{f} g_\pm(k)  + \frac{1}{4}\left( \frac{f'}{f}\right)^2  + \frac{2u}{f} \left[ -5 Q^2  u+ 3(1 - Q^2)  \right]   \label{eq:37}
\end{align}
We shall call the $\Psi_\pm(u | \mathfrak{q})$ as the modified master fields in subsequent discussions.

\subsection{Transverse Polarization on Boundary Theory}
\label{sec:transv-corr-bound}

The quantum effective action of the boundary field theory corresponds to the bulk action (\ref{eq:17}) evaluated at (\ref{eq:22}), with $h_{\mu\nu}$ and
$a_\mu$ the solutions of eqs. (\ref{eq:25})-(\ref{eq:28}) and consists of purely boundary terms to the quadratic orders in $h_{\mu\nu}$ and $a_\mu$. The on-shell action is therefore 
a quadratic functional of the boundary 
values of $h^y_t$ and $a_y$ with the coefficients corresponding to various two point 1PI Green's functions in strong coupling. In particular, the coefficient of $a_y^2$ gives rise to the 
transverse component of the polarization tensor whose weak coupling limit is considered in section 2. With the boundary condition $h_{\mu\nu}=0$ at $z=0$, the on-shell action
contains only $a_y^2$ term and reads   
\begin{align}
  \big( \mathcal{S}_\text{eff} \big)_\text{EM} =   K \int \ \mathrm{d}^3x \, \bigg[ \sqrt{- \bar g} \bar g^{uu}  \,  \bar g^{yy} \, a_y' \, a_y  \bigg] \Bigg|_{u  \to 0}   \label{eq:39}
\end{align}
The Fourier component of the  transverse polarization tensor is therefore  given by
\begin{align}
  \mathcal{C}_{yy} (\mathfrak{w},\mathfrak{q}) &= K\, \lim_{u \to 0}\sqrt{- \bar g} \, \bar g^{uu} \bar g^{yy} \; \frac{a_y'}{a_y}      \\
 &=K \, \lim_{u \to 0} f\, \big[ \log a_y\big]' \label{eq:41}
\end{align}
with $a_y$ the solution of (\ref{eq:25})-(\ref{eq:28}) subject to the condition that $h_{\mu\nu}=0$ at the boundary. In the static case, we write
$\mathcal{C}_{yy}(0,\mathfrak{q})\equiv\sigma^{\mathrm{tr}}(\mathfrak{q})$ and the equations to be solved for $h_{\mu\nu}$ and $a_\mu$ are (\ref{eq:29}) and (\ref{eq:30}).

It follows from the definition of the master fields (\ref{eq:31}) and the transformation (\ref{eq:35}) that 
\begin{align}
  a_y &= - \frac{\Phi_+ - \Phi_-}{2 Q^2 (g_+ - g_-)}  \label{eq:42}  \\
  &= \frac{1}{4 Q^2}\frac{ \big(\Psi_- -\Psi_+ \big) }{\sqrt{ f (Z^2+  \mathfrak{ q}^2) } }  \label{eq:43}
\end{align}
here we denote
\begin{align}
  Z :=  \frac{3}{4}\left(1 + \frac{1}{Q^2} \right)   \label{eq:44}
\end{align}
Substituting (\ref{eq:43}) into (\ref{eq:41}), we find that
\begin{align}
\sigma^{\mathrm{tr}}(k)=\mathcal{C}_{yy} (0, \mathfrak{q})= K\, \lim_{u \to 0} f \, \bigg[ \log \big( \Psi_- - \Psi_+\big)\bigg]'   \label{eq:45}
\end{align}

\subsection{The analyticity of the solution with respect to the momentum}
\label{sec:analyt-solut-with}

For the analyticity of the solution to the master equation (\ref{eq:32}) with respect to the momentum $\mathfrak{q}$, it is important to notice that 1)
the coefficient in front of $\Phi_\pm$ in (\ref{eq:32}) has no double poles with respect to $1-u$ and is a polynomial in the modified momentum $k$ defined in (\ref{eq:47}),
2) the coefficient of $\Phi_\pm^\prime$ is independent of $k$ and 3) the boundary $u=0$ is an ordinary point of the equation.

Let us consider the power series solution in the neighborhood of the regular point at the horizon, $u=1$, where both indices of the solution are zero. We shall observe that , 
Because of the first two points mentioned above, the denominator of  each coefficients in this series solution do not involve the modified momentum $k$, which manifests the analyticity of $\Phi_\pm$ with respect to $k$.
In terms of the new variable $\zeta = 1 - u$, the master equation (\ref{eq:32}) can be written in the form
\begin{equation}
  \zeta \Phi^{\prime\prime}+p(\zeta) \Phi^\prime+q(\zeta) \Phi=0
\end{equation}
where $p(\zeta)$ and $q(\zeta)$ are rational functions with three simple poles at roots of the cubic polynomial
$f(1-\zeta)/\zeta$, which are located outside the segment of real axis, $0 \le \zeta \le 1$. The subscript $\pm$ pertaining to the master field $\Phi$ have been suppressed. 
Upon Taylor expansions
\begin{equation}
  p(\zeta)=\sum_{l=0}^\infty a_l \zeta^l
\end{equation}
and
\begin{equation}
  q(\zeta)=\sum_{l=0}^\infty b_l \zeta^l
\end{equation}
the Frobenius method formulates the solution analytic at $\zeta=0$ by the power series
\begin{equation}
  \Phi=\sum_{n=0}^\infty c_n\zeta^n   \label{frobenius}
\end{equation}
with the coefficients determined recursively by
\begin{equation}
  c_n=-\frac{1}{n}\sum_{l=1}^n\left[a_l(n - l)+b_{l-1}\right]c_{n-l}.
\end{equation}
 Given the condition $c_0=1$, $c_n$ is a polynomial in $k$ and is thereby analytic in $k$. It follows from the Weierstrass theorem that the solution (\ref{frobenius}) 
and its derivative with respect to $u$ are all analytic in $k$ within the radius of convergence of $\zeta$. The radius of convergence, $r$, of the series (\ref{frobenius}) is 
the closest distance from the horizon $\zeta=0$ to other regular points of the equation. If $r>1$, our statement of the analyticity extends readily to the 
solution at the boundary $\zeta=1$. Otherwise, we may choose a $\zeta=\zeta_0$ such that $0 < \zeta_0 < r$ and its distance to the nearest regular point is  $d > r- \zeta_0$ .

Upon Taylor expansions
\begin{equation}
  \frac{p(\zeta)}{\zeta}=\sum_{n=0}^\infty\alpha_n(\zeta-\zeta_0)^n
\end{equation}
and
\begin{equation}
  \frac{q(\zeta)}{\zeta}=\sum_{n=0}^\infty\beta_n ( \zeta - \zeta_0)^n
\end{equation}
we obtain the series solution in the neighborhood of the ordinary point $\zeta_0$,
\begin{equation}
  \Phi=\sum_{n=0}^\infty\gamma_n  (\zeta - \zeta_0)^n     \label{ordinary}
\end{equation}
where the coefficients determined recursively by
\begin{equation}
  \gamma_{n+2}=-\frac{1}{(n+1)(n+2)}\sum_{l=0}^n\left[(n- l + 1)\alpha_k\gamma_{n - l + 1}+\beta_l \gamma_{n - l}\right] \label{ordinary_1}
\end{equation}
with $\gamma_0$ and $\gamma_1$ given by the series solution (\ref{ordinary}) and its derivative evaluated at $\zeta=\zeta_0$.
All components on RHS of (\ref{frobenius}) are analytic in $k$, so is the series solution (\ref{ordinary}) and the validity of our 
analyticity statement has been extended to $\zeta < \zeta_0 + d$, beyond the original radius of convergence. This process may be repeated if necessary
until we reach the boundary. It follows from (\ref{eq:42}) and (\ref{eq:45}) that $C_{yy}$ is a ratio of two analytic functions and is thereby
a meromorphic function of the modified momentum $k$. 

Substituting $k=\sqrt{\mathfrak{q}^2 + \left[ \frac{3}{4}\left( 1 + \frac{1}{Q^2}\right)\right]^2}$, we note that the expression of $C_{yy}$ 
of (\ref{eq:45}) is symmetric with respect to the sign of the square root and the corresponding branch cut is canceled in the combination. This 
concludes the proof that $C_{yy}$ is a meromorphic function of the true momentum.

\section{ The WKB Approximation at Large Momentum Magnitude }
\label{sec:wkb-appr-analyt}
  
\subsection{The WKB solution}
\label{sec:large-imaginary-part}

In the last subsection, we derived an holographic relation between the modified master fields, $\Psi_\pm$,  and the transverse component of the polarization tensor, 
$\sigma^{\mathrm{tr}}(q)$. The singularities of $\sigma^{\mathrm{tr}}(q)$ on the complex $q$-plane corresponds to the nontrivial solutions for $a_y$ that vanish at the 
boundary. While the singularities near the real $q$-axis can only be accessed numerically, the WKB approximation applied to the Schr\"{o}dinger-like equation (\ref{eq:36})
enables us to locate explicitly the singularities far from the real axis, thereby offers an analytic proof of the existence of Friedel-like singularities of the 
holographc polarization tensor. 

Let us write the modified momentum in terms of its real and imaginary parts
\begin{align}
  k = w + \mathrm{i}p  \label{eq:51}
\end{align}
and consider the case that $p \gg w$. The ``potential'' of the Schr\"{o}dinger-like equation (\ref{eq:37}) can be approximated as  
\begin{align}
  V_\pm (u | q)&\simeq -\frac{Q^2}{f}\, (k \pm u)^2  =  \frac{Q^2}{f}\big[ p - \mathrm{i}(w \pm u) \big]^2  \label{eq:49}
\end{align}
and the  WKB solution reads
\begin{align}
  \big(\Psi_\pm\big)_{_\text{WKB}} &= \frac{f^{1/4}}{Q^{1/2}} \bigg( C_1^\pm \, \exp \bigg\{ - \mathrm{i} \int_u^1 \ \mathrm{d}v \, \frac{Q}{\sqrt{f}} \big[ p - \mathrm{i}(w \pm v)  \big] \bigg\} \notag \\
 &\hspace{4cm} + C_2^\pm \exp\bigg\{+ \mathrm{i}  \int_u^1 \ \mathrm{d}v \, \frac{Q}{\sqrt{f}} \big[ p - \mathrm{i}(w \pm v)  \big] \bigg\}  \bigg)  \label{eq:52}
\end{align}
with $C_1^\pm$ and $C_2^\pm$ constants to be determined. The difference between (\ref{eq:52}) and the exact solution is of the order $\mathcal{O}(p^{-1})$ and can be ignored for a large $p$. 

\subsubsection*{Matching the WKB solution with the near  horizon solution}
\label{sec:match-betw-asympt}
     
Near the horizon $u \sim 1$, we have 
\begin{align}
f(u) \simeq (3 -Q^2)(1 - u)
\end{align}
thus,
\begin{align}
  \mathrm{i}  \int_u^1 \ \mathrm{d}u \, \frac{Q}{\sqrt{f}} \big[ p - \mathrm{i}(w \pm u)  \big] \simeq  2\lambda_\pm\sqrt{1- u} 
\end{align}
where 
\begin{align}
  \lambda_\pm := \frac{Q[k \pm 1]}{\sqrt{3 - Q^2}}  \label{eq:54}
\end{align}
The WKB solution can be extended to this region as long as $\lambda_\pm\sqrt{1- u} \gg 1$ following from the condition of the approximation $|V_\pm^\prime| \ll |V_\pm|^{\frac{3}{2}}$.
and the asymptotic form there is given by
\begin{align}
  \big(\Psi_\pm\big)_{_\text{WKB}} =  \frac{(3 - Q^2)^{1/4}}{Q^{1/2}} [1-u]^{1/4} \cdot \bigg( C_1^\pm \, e^{ -2 \sqrt{\lambda_\pm}\, \sqrt{1-u}} + C_2^\pm \, e^{ + 2 \sqrt{\lambda_\pm}\, \sqrt{1-u}} \bigg)  \label{eq:53}
\end{align}

On the other hand, the modified master field equations (\ref{eq:36}), in the horizon limit,  can be transformed into a modified Bessel equation of the zeroth order  with the general solution
\begin{align}
  \Psi_\pm = \sqrt{1-u} \cdot \bigg[  b_\pm \, I_0(2 \sqrt{\lambda_\pm} \sqrt{1-u}) + c_\pm\,  K_0(2 \sqrt{ \lambda_\pm} \sqrt{1-u}) \bigg]   \label{eq:55}  
\end{align}
with $b_\pm$ and $c_\pm$ the constant coefficients.

The function $K_0(2 \sqrt{\lambda_\pm} \sqrt{1-u})$ diverges logarithmically as $u\to 1$, i.e.
\begin{align}
 K_0( u \to 1) \sim - \log\left( \sqrt{\lambda_\pm} \sqrt{1-u}\right)   \label{eq:56}
\end{align}
From (\ref{eq:43}) and (\ref{eq:55})  , as  $u \to 1$  , we have 
\begin{align}
  a_y \sim K_0  \sim (c_+-c_-)\log \left( \sqrt{\lambda_\pm} \sqrt{1-u } \right)
\end{align}
which would make the on-shell action divergent and should be dropped by setting $c_+-c_-=0$. Now we show that the absence of curvature singularity at the horizon requires that 
$c_++c_-=0$ and consequently, $c_- = c_+ =0$.
Following from (\ref{eq:31}) and (\ref{eq:46}),  we get the relation
\begin{equation}
\Phi_++\Phi_-=-\frac{2\mu}{u}{h^y_t}^\prime+[8Q^2u-3(1+Q^2)]a_y. \label{PhiPhi}
\end{equation}
The LHS of (\ref{PhiPhi}) has a logarithmic divergence with the coefficient $c_- + c_+$, which indicates a logarithmic singularity in ${h_t^y}'$. 
The logarithmic singularity in 
${h^y_t}^\prime$ will cause one of the component of the Riemann tensor diverges like $\frac{1}{1-u}$ at the horizon and the invariant $R_{\mu\nu\rho\sigma}R^{\mu\nu\rho\sigma}$ diverges like $\frac{1}{(1-u)^2}$ there.
Therefore  both $c_\pm$ should vanish and only $I_0$ function in (\ref{eq:55}) is left over. 
  
In the region $\lambda_\pm\sqrt{1- u} \gg 1$, where both the WKB solution and the Bessel function solution approximate, 
we have two forms of the same solution, (\ref{eq:53}) and the asymptotic form of(\ref{eq:55}) at $c_\pm=0$, i.e.
\begin{align}
  \Psi_\pm =  \frac{b_\pm}{2\sqrt{\pi}}\lambda_\pm^{-\frac{1}{4}}(1-u)^{\frac{1}{4}}
\bigg(\mathrm{i} e^{ -2 \sqrt{\lambda_\pm}\, \sqrt{1-u}} + e^{ + 2 \sqrt{\lambda_\pm}\, \sqrt{1-u}} \bigg)  \label{eq:55_1}
\end{align}
where we have substituted the asymptotic form of $I_0(z)$ for large $|z|$,
\begin{align}
  I_0(z) \simeq \frac{1}{\sqrt{2 \pi z}} \bigg[ \mathrm{i} \, e^{-z}  + e^z   \bigg]   \label{eq:57}
\end{align}
Note that for the limit of $z\to\infty$ along a line parallel to the imaginary axis, both terms inside the bracket of (\ref{eq:57}) have to be retained.
Matching (\ref{eq:53}) and (\ref{eq:55_1}), we obtain that
\begin{align}
  C_1^\pm = \mathrm{i} C_2^\pm := \mathrm{i} \, C_\pm \label{eq:59}
\end{align}

\subsection{Constraint on fluctuations of metric field on the AdS boundary}
\label{sec:constr-fluct-metr}

Carrying the WKB solution (\ref{eq:52}) with the coefficients given by (\ref{eq:59}) to the AdS boundary,  we obtain

\begin{align}
  \big(\Psi_\pm\big)_{_\text{WKB}}  \bigg|_{u \to 0} = \frac{C_\pm}{Q^{1/2}} \bigg[ \,  \mathrm{i} \exp\big\{- Q\left[ k L_1 + (\pm) L_2\right] \big\} +  \exp\big\{ Q[ k L_1 + (\pm) L_2 ] \big\}\bigg]  \label{eq:68}
\end{align}
where $L_1$ and $L_2$ are two elliptic integrals given by
\begin{equation}
\begin{aligned}
  L_1 &:=  \int_0^1 \frac{1}{ \sqrt{f(v)} } \ \mathrm{d}v   \\
 L_2 &:=  \int_0^1 \frac{v}{ \sqrt {f(v)} } \ \mathrm{d}v  \label{eq:67}
\end{aligned}
\end{equation}
The relationship between $C_+$ and $C_-$ remains to be fixed in order to extract $C_{yy}$ and the boundary condition $h^y_t=0$ at $u=0$ serves the purpose. 

The equation for the metric fluctuation, (\ref{eq:30}), near the boundary takes the asymptotic form 
\begin{align}
  {h_t^y}'' - \frac{2}{u}\, {h_t^y}' - Q^2 \, \mathfrak{q}^2\, {h_t^y} = \frac{4 Q^2}{\mu} u^2 \, a_y'  \label{eq:61-0}
\end{align}
Leaving out the inhomogeneous term on RHS, the homogeneous equation 
\begin{align}
  {h_t^y}'' - \frac{2}{u}\, {h_t^y}' - Q^2 \, \mathfrak{q}^2\, {h_t^y} = 0  \label{eq:61}
\end{align}
hence at the regular point $u=0$ , we have  
\begin{align}
  h_t^y \sim 1 \hbox{ or } h_t^y \sim u^3  \label{eq:62}  
\end{align}
Because $a_y$ is nonsingular at $u=0$, restoring the inhomogeneous term will modify neither asymptotic behavior in (\ref{eq:62}). The boundary condition $h^y_t=0$ 
eliminates the first one and we are left with
\begin{align}
  h_t^y = O\big(  u ^3 \big) \label{eq:63}
\end{align}

Owing to (\ref{eq:31})(\ref{eq:46}) ,and (\ref{eq:44}), we have
\begin{align}
  {h_t^y}' = O\big( u^2 \big)= -\frac{u}{2 \mu} \cdot \Big( \left[ 1 + \frac{Z }{k} \right] \, \Phi_- + \left[ 1 - \frac{Z }{k}  \right] \, \Phi_+  \Big) \bigg|_{u=0} + O\big( u^2 \big)   \label{eq:64}
\end{align}
which implies a condition for $C_\pm$,  (for $p \ne 0$)
\begin{align}
0 = \big[ k + Z \big] \,   \Psi_-  + \big[k - Z \big] \, \Psi_+  \label{eq:65}
\end{align}
Substituting into (\ref{eq:68}) and using the notations in (\ref{eq:67}) , we obtain
\begin{align}
  0 = (k + Z) \big[ \mathrm{i} + e^{ 2 Q( k L_1 - L_2)}\big] \cdot C_- + (k - Z) \big[  \mathrm{i} \, e^{ - 2 Q L_2} + e^{ 2 Q k L_1 }  \big] \cdot C_+  \label{eq:66}
\end{align}

\subsection{The asymptotic singularities on the complex momentum plane}
\label{sec:analyt-struct-transv}

To explore the analyticity of the transverse polarization, we look for the singularities of $\mathcal{C}_{yy}(\mathfrak{q})$  on the $q$-complex plane. 
From (\ref{eq:45}) and (\ref{eq:68}) we find the explicit formula for the photon self-energy at large momentum magnitude,
\begin{align}
  \mathcal{C}_{yy} (q) &= K\, Q\, k \cdot \Big( 1 -  \frac{ 2 \cdot \big[ e^{ 2 Q(k L_1 - L_2)} \cdot C_- - e ^{ 2 Q k L_1} \, C_+  \big] }{ \big[ \mathrm{i} + e^{ 2 Q( k L_1 - L_2)} \big] \cdot C_- - \big[ \mathrm{i} \, e^{ - 2 Q L_2} + e^{ 2 Q k L_1 } \big] \cdot C_+ }  \Big)   \label{eq:69}
\end{align}
the zeros of the denominator implies the singularities of $\mathcal{C}_{yy}$, i.e.
\begin{align}
  0 = \big[ \mathrm{i} + e^{ 2 Q( k L_1 - L_2)} \big] \cdot C_- - \big[ \mathrm{i} \, e^{ - 2 Q L_2} + e^{ 2 Q k L_1 } \big] \cdot C_+  \label{eq:70}
\end{align}

At this point, we are sufficiently equiped to tackle the problem, together with (\ref{eq:66}), the existence of nontrivial solutions for $C_\pm$ implies that 

\begin{equation}
\begin{aligned} 0 &=
  \begin{vmatrix}
    (k + Z) \cdot \big[ \mathrm{i} + e^{ 2 Q( k L_1 - L_2)} \big] \quad &   (k -Z) \cdot \big[ \mathrm{i} \, e^{ - 2 Q L_2} + e^{ 2 Q k L_1 } \big]  \\
    \big[ \mathrm{i} + e^{ 2 Q( k L_1 - L_2)} \big]   & \quad - \big[ \mathrm{i} \, e^{ - 2 Q L_2} + e^{ 2 Q k L_1 } \big]
  \end{vmatrix} \\
  &= (- 2 k) \cdot \big[ \mathrm{i} + e^{ 2 Q( k L_1 - L_2)} \big] \cdot \big[ \mathrm{i} \, e^{ - 2 Q L_2} + e^{ 2 Q k L_1 } \big]
  \label{eq:76}
\end{aligned}
\end{equation}
Since $k \ne 0$, the locations of the poles are given by:
\begin{equation}
\big[ \mathrm{i} + e^{ 2 Q( k L_1 - L_2)} \big] \cdot \big[ \mathrm{i} \, e^{ - 2 Q L_2} + e^{ 2 Q k L_1 } \big]
\end{equation}

1. If $C_-  \ne 0$ ,by (\ref{eq:76}),  we have
\begin{equation}
\begin{aligned}
  0 &= \mathrm{i} + e^{ 2 Q( k L_1 - L_2)}     \\
 &= e^{ 2 Q(L_1 w - L_2) } \cos \big( 2 Q L_1\, p\big) + \mathrm{i} \left[  1 + e^{ 2 Q (L_1 w -  L_2 ) }\sin \big( 2 Q L_1\, p\big)\right]
\end{aligned}
\end{equation}
Consequently,
\begin{align}
  w &= \frac{L_2}{L_1} \label{eq:73} \\
  p &= \frac{\pi}{2 Q L_1 } \left(2n - \frac{1}{2} \right)  ,  \qquad n \in \mathbb{Z}   \label{eq:74}
\end{align}

2. If $C_+  \ne 0$ , by (\ref{eq:76})
\begin{equation}
\begin{aligned}
  0 &=  \mathrm{i} \, e^{ - 2 Q L_2} + e^{ 2 Q k L_1 } \\
 &= e^{2 Q L_1 \, w } \cos \big( 2 Q L_1\, p\big)  + \mathrm{i} \left[ e^{- 2Q L_2} + e^{ 2 Q L_1 \, w } \sin \big( 2 Q L_1\, p\big)  \right] 
 \label{eq:75}
\end{aligned}
\end{equation}
Consequently, 
  \begin{align}
    w &= - \frac{L_2}{L_1}   \label{eq:3}\\
    p &= \frac{\pi}{2 Q L_1} \left( 2\, n - \frac{1}{2} \right) , \qquad n \in \mathbb{Z}  \label{eq:6}
  \end{align}

Therefore, in the regime far away from the real axis,   all isolated poles $k = w + \mathrm{i}p$ are located on the two lines parallel to the imaginary axis, with the  real parts dependent on the 
chemical potential as expected.

As the tempearture $T\to 0$\footnote{For a holographic superconductor, the normal phase described by the action (\ref{eq:17}) at $T=0$ is unstable against onset of the 
long range order described by a complex scalar field. Alternatively, we may consider the action (\ref{eq:17}) without the complex scalar field being a gravity dual of some 
strongly coupled QED3.}, the two elliptic integrals $L_1$ and $L_2$ in (\ref{eq:67}) can be approximated analytically and we have
\begin{equation}
\begin{aligned}
  L_1 \simeq  \frac{1}{\sqrt{6}}\ln\frac{72\mu}{(2\sqrt{3}+3\sqrt{2})\pi T}  \\
 L_2 \simeq  \frac{1}{\sqrt{6}}\ln\frac{72\mu}{(2\sqrt{3}+3\sqrt{2})\pi T}-\frac{1}{\sqrt{3}} \ln\frac{4+3\sqrt{2}}{1+\sqrt{3}}. \label{elliptic}
\end{aligned}
\end{equation}
Consequently, the poles merge along the two lines parallel to the imaginary axis of the complex $q$-plane for large $p$ with the real part $w\to \pm1$. 
The rate of merging ($\sim (\ln\frac{\mu}{T})^{-1}$) is, however, much slower than the one-loop case ($\sim T/\mu$). In the high tempearture limit, $Q\to 0$ ,  the distance between successive poles along the asymptotes approaches a finite limit instead of growing linearly with 
the $T$ as the one-loop result. These difference may be attributed to the strong couling of the boundary field theory described by the holography. 
It is worth mentioning that the criterion of the WKB solution, $Qq \gg 1$ as is implied by (\ref{eq:37}) , pushes the domain on the complex  $q$ -plane where the asymptotic distribution of the poles, (\ref{eq:73}) , (\ref{eq:74}) , (\ref{eq:3})  and (\ref{eq:6})approximates further away from the real axis with increasing  $T$  (decreasing  $Q$ ).

\section{Discussions and Outlooks}
\label{sec:disc-concl}

Let us recapitulate what we have done in this paper. We started with a field theoretic calculation of the transverse component of the
polarization tensor in weak coupling and located its singularities. They all lies on two lines parallel to the imaginary axis with the real part equal to
$\pm \mu$. Then we moved to the normal phase of the holographic superconductivity beyond the probe limit and extracted the
transverse component of the polarization tensor in terms of the solution of the linearized Einstein-Maxwell equations. We were
able to prove that the static and transverse polarization tensor is a meromorphic function of momentum. At large momentum magnitude
where the WKB approximation works, we found equally spaced singularities along two lines parallel to the imaginary axis with the real parts given by (\ref{eq:73}) and (\ref{eq:3}), 
which are chemical potential dependent, mimicking the singularity distribution of the polarization tensor in weak coupling. The same argument also ruled out
the complex singularities elsewhere at large momentum magnitude. The natures of the singularities,
however are different. They are branch points in weak coupling and poles in holography. Mathematically it is possible that,
the isolated poles merge into a branch cut in weak coupling, as is evident in the example below
\begin{equation}
  \sum_{n=1}^\infty \frac{\sqrt{\frac{\epsilon}{n}}}{z-z_0+n\epsilon}\simeq \int_0^\infty dx\frac{x^{-\frac{1}{2}}}{z-z_0+x}=\frac{\pi}{\sqrt{z-z_0}}   \label{eq:9}
\end{equation}
for $\epsilon \ll 1$ , where the infinite sum turns into an integral.

For a complex momentum not far from the real axis, the WKB approximation failed and we were unable to gain more insight than the
numerical work reported in \cite{Blake2015}. But nevertheless, the explicit WKB solution provided for the first time an analytic evidence of the Friedel-like
singularities in the holographically implied Green functions which makes the conjectured gauge-gravity duality  more plausible. 

The Friedel-like singularities pertaining to the $\mathcal{ C}_{tt}$ component of the polarization tensor, which belongs to the even parity group under $y \rightarrow -y$ is 
being currently investigated analytically \cite{YinLei}. Dispite of the technical complexity with the even parity, our preliminary result suggests that the asymptotic distribution of the complex 
momentum poles far from the real axis is identical to that of $\mathcal{ C}_{yy}$ explored in this paper. As to the properties of the poles close to the real axis, their 
migration to the imaginary axis with increasing temperature and their gaps to the real axis even at zero temperature (exponetial decay of the charge fluctuation at zero 
temperature), observed numerically in Ref.\cite{Blake2015} are beyond the WKB approximation employed here.

\appendix

\section{An identity employed in the weakly-coupled system}
\label{sec:an-identity-employed}

We consider a contour integral 
\begin{align}
  \mathcal{I}_1 = \oint_{\mathcal{C}} \frac{\mathrm{d}z}{2\pi \mathrm{i}} \, \frac{1}{e^{\beta z} + 1} \left( \frac{1}{z + \mu - p } - \frac{1}{z + \mu + p} \right)
\end{align}
which has two kinds of simple poles for its integrand, one is a series of poles on the imaginary axis: $z = \mathrm{i}\nu /\beta$ , the other is two on real axis: $z_1 = -(\mu - p)$ and $z_2 = -(\mu + p)$ . We have two ways to deform the contour and get the result in a different form 
\begin{equation}
\begin{aligned}
 \mathcal{I}_1 &=  T \sum_\nu \left( \frac{1}{\mathrm{i} \nu + \mu - p } - \frac{1}{\mathrm{i} \nu + \mu + p}\right) \\
&= \frac{1}{e^{\beta (p - \mu)} +1} - \frac{1}{e^{-\beta(p + \mu)} +1}
\end{aligned}
\end{equation}
rearranging the last formula, we obtain the identity
\begin{align}
  \frac{1}{e^{\beta (p - \mu)} +1} + \frac{1}{e^{\beta(p + \mu)} +1} = 1 + T \sum_\nu \left( \frac{1}{\mathrm{i} \nu + \mu - p } - \frac{1}{\mathrm{i} \nu + \mu + p}\right)  \label{eq:8}
\end{align}
This identity is employed in (\ref{eq:7}).


\subsection*{Acknowledgments}
\label{sec:acknowledgments}

L.Y. and T.K.L. acknowledge the  support  by  MOST 104-2112-M-001-005 ,  D. F. H. and H.C. R. are supported by by the Ministry of Science and Technology of China (MSTC) under the “973” Project No.  2015CB856904. And L.Y. also supported by QLPL2015P01 under No. 201508;  D. F. H. and H.C. R.  by NSFC under Grants No.  11375070, No.  11221504 and No. 11135011. 

\bibliographystyle{JHEP}

\begin{thebibliography}{10}

\bibitem{Witten1998b}
  E.~Witten, \emph{{Anti-de Sitter space and holography}},
  \href{http://dx.doi.org/10.1186/1753-6561-3-S7-S29}{\emph{Adv.Theor.Math.Phys.}
    {\bf 2} (1998) 253--291}, [\href{http://arxiv.org/abs/9802150}{{\tt
      9802150}}].

\bibitem{Maldacena1998}
  J.~Maldacena, \emph{{The large N Limit of superconformal field theories and
      supergravity}},
  \href{http://dx.doi.org/10.1023/A:1026654312961}{\emph{Advances in
      Theoretical and Mathematical Physics} {\bf 2} (1998) 231--252},
  [\href{http://arxiv.org/abs/9711200}{{\tt 9711200}}].

\bibitem{Aharony1999}
  O.~Aharony, S.~Gubser, J.~Maldacena, H.~Ooguri and Y.~Oz, \emph{{Large N Field
      Theories, String Theory and Gravity}},
  \href{http://dx.doi.org/10.1016/S0370-1573(99)00083-6}{\emph{Physics Reports}
    {\bf 323} (1999) 183--386}, [\href{http://arxiv.org/abs/9905111v3}{{\tt
      9905111v3}}].

\bibitem{Klebanov1999}
  I.~R. Klebanov and E.~Witten, \emph{{AdS/CFT Correspondence and Symmetry
      Breaking}},  \href{http://arxiv.org/abs/9905104}{{\tt 9905104}}.

\bibitem{Maldacena1998b}
  J.~Maldacena, \emph{{Wilson Loops in Large N Field Theories}},
  \href{http://dx.doi.org/10.1103/PhysRevLett.80.4859}{\emph{Physical Review
      Letters} {\bf 80} (1998) 4859--4862},
  [\href{http://arxiv.org/abs/9803002}{{\tt 9803002}}].

\bibitem{Policastro2001}
  G.~Policastro, D.~T. Son and A.~O. Starinets, \emph{{Shear Viscosity of
      Strongly CoupledN 5 4 Supersymmetric Yang-Mills Plasma}},
  \href{http://dx.doi.org/10.1103/PhysRevLett.87.081601}{\emph{Physical Review
      Letters} {\bf 87} (2001) 081601}.

\bibitem{Gubser2008}
  S.~S. Gubser, \emph{{Breaking an Abelian gauge symmetry near a black hole
      horizon}}, \href{http://dx.doi.org/10.1103/PhysRevD.78.065034}{\emph{Physical
      Review D - Particles, Fields, Gravitation and Cosmology} {\bf 78} (2008) },
  [\href{http://arxiv.org/abs/0801.2977}{{\tt 0801.2977}}].

\bibitem{Hartnoll2008}
  S.~A. Hartnoll, C.~P. Herzog and G.~T. Horowitz, \emph{{Building a holographic
      superconductor}},
  \href{http://dx.doi.org/10.1103/PhysRevLett.101.031601}{\emph{Physical Review
      Letters} {\bf 101} (2008) 1--4}, [\href{http://arxiv.org/abs/0803.3295}{{\tt
      0803.3295}}].

\bibitem{Hartnoll2008a}
  S.~A. Hartnoll, C.~P. Herzog and G.~T. Horowitz, \emph{{Holographic
      Superconductors}},  \href{http://arxiv.org/abs/0810.1563}{{\tt 0810.1563}}.

\bibitem{Son2002}
  D.~T. Son and a.~O. Starinets, \emph{{Minkowski-space correlators in AdS/CFT correspondence: recipe and applications}},
  \href{http://arxiv.org/abs/0205051}{{\tt 0205051}}.

\bibitem{Policastro2002}
  G.~Policastro, D.~T. Son and a.~O. Starinets, \emph{{From AdS/CFT
      correspondence to hydrodynamics. II. Sound waves}},
  \href{http://arxiv.org/abs/0210220}{{\tt 0210220}}.

\bibitem{Hou2010}
  D.~F. Hou, J.~R. Li, H.~Liu and H.~C. Ren, \emph{{The momentum analyticity of
      two point correlators from perturbation theory and AdS/CFT}},
  \href{http://dx.doi.org/10.1007/JHEP07(2010)042}{\emph{Journal of High Energy
      Physics} {\bf 2010} (2010) }, [\href{http://arxiv.org/abs/1003.5462}{{\tt
      1003.5462}}].

\bibitem{Kohn1965}
  W.~Kohn and J.~M. Luttinger, \emph{{New Mechanism for Superconductivity}},
  \href{http://dx.doi.org/10.1103/PhysRevLett.15.524}{\emph{Phys. Rev. L.} {\bf
      15} (1965) 63--471}.

\bibitem{Galitski2003}
  V.~Galitski and S.~{Das Sarma}, \emph{{Kohn-Luttinger pseudopairing in a
      two-dimensional Fermi liquid}},
  \href{http://dx.doi.org/10.1103/PhysRevB.67.144520}{\emph{Physical Review B}
    {\bf 67} (2003) 144520}, [\href{http://arxiv.org/abs/0211355v2}{{\tt
      0211355v2}}].

\bibitem{Egger1995}
  R.~Egger and H.~Grabert, \emph{{Friedel oscillations for interacting fermions
      in one dimension}},  \href{http://arxiv.org/abs/9509100}{{\tt 9509100}}.

\bibitem{Blake2015}
  M.~Blake, A.~Donos and D.~Tong, \emph{{Holographic charge oscillations}},
  \href{http://dx.doi.org/10.1007/JHEP04(2015)019}{\emph{Journal of High Energy
      Physics} {\bf 2015} (2015) }.

\bibitem{Edalati2010c}
  M.~Edalati, J.~I. Jottar and R.~G. Leigh, \emph{{Shear modes, criticality and extremal black holes}},
  \href{http://dx.doi.org/10.1007/JHEP04(2010)075}{\emph{Journal of High Energy
      Physics} {\bf 2010} (2010) }, [\href{http://arxiv.org/abs/1001.0779}{{\tt 1001.0779}}].

\bibitem{Kodama2004}
  H.~Kodama and A.~Ishibashi, \emph{{Master Equations for Perturbations of
      Generalised Static Black Holes with Charge in Higher Dimensions}},
  \href{http://dx.doi.org/10.1143/PTP.111.29}{\emph{Progress of Theoretical
      Physics} {\bf 111} (2004) 29--73}, [\href{http://arxiv.org/abs/0308128}{{\tt
      0308128}}].

\bibitem{YinLei} 
  L. Yin, H. C. Ren, T. K. Lee and D. F. Hou, \emph{{Momentum analyticity of the electric polarizability of  a holographic QED$_3$  }},work in progress.

\end{thebibliography}

\providecommand{\href}[2]{#2}\begingroup\raggedright\endgroup

\end{document}